# Written by AI, Managed by AI: Semantic Space Control and Index Sickness Elimination Across 391 Consecutive Sessions

**Hui Zhang (张慧)** Shenzhen Yunxi Technology Co., Ltd. zhanghui@ecloudriver.com

**Shuren Song (宋树仁)** Information Technology Center, Tsinghua University songsr@tsinghua.edu.cn

---

## Abstract

The prevailing engineering intuition for addressing conceptual drift in long-horizon LLM collaboration is to trade more formal constraints for more reliable outputs—designing symbolic identifier systems for task entities, accumulating defensive rules in System Prompts, expanding context windows to accommodate ever more complete constraint indices. Our engineering record shows that in long-horizon settings, this direction may produce effects contrary to design intent.

Using action research methods in a real software refactoring project (Bang-v3) spanning approximately one month and 391 collaborative sessions, we document and analyze the failure process of the above strategies in this project. When the symbolic system exceeds a threshold of complexity, LLMs do not become more accurate with more precise constraints—instead, they abandon genuine understanding of business semantics, retreat to self-referential reasoning within the symbolic layer, and generate outputs that appear internally consistent but are physically disconnected from reality. We name this failure pattern **"Index Sickness"**, and its canonical manifestation **"Phantom Legislation."**

This failure has a counterintuitive interpretation: the core obstacle to long-horizon LLM collaboration is not insufficient AI memory, but the presence of historical noise in the active context that should not be there. Recognizing this, the solution becomes simple—it is at its core a return to an overlooked common truth. We name this truth the **"Pang Principle (Semantic Vitality Law)"**: in human–AI collaboration, natural language carrying explicit purpose conveys far greater information quality than symbolic expression. From this recognition, we design and validate its physical engineering mechanism: **"Baseline-Log Physical Separation."** In the same project, this mechanism reduced AI Instructions volume by ~75%, and across the subsequent ~150 sessions, no recurrence of Index Sickness was observed; Owner correction density dropped by 33%, and the 9 confirmed Index Sickness cases identified pre-upgrade fell to zero post-upgrade.

---

# I. Introduction

## 1.1 Problem Background and Paper Perspective

In recent years, the AI-assisted software engineering field has produced a body of improvements all pointing in the same direction: more precise symbolic identifier systems and Prompt engineering formalization [1][2], heavier structured-memory architectures [3][4][5], larger context windows and retrieval-augmented injection. These works all answer the same question—**how to provide AI with better inputs.** We collectively call this class of approaches **Supply-side** improvements: their shared assumption is that AI output quality is determined by the completeness and determinism of supply-side inputs—the more precise and complete the input, the more reliable the output.

These improvement paths have in recent years been consolidated by the industry into a unified practice direction, which Anthropic terms **"Context Engineering"** [6]: "curating and maintaining the optimal set of tokens from a constantly evolving universe of possible information at each inference step, to maximize the likelihood of the desired outcome"—covering system prompt design, retrieval-augmented injection, long-horizon compaction, and structured memory, all unified by the demand to manage the model's limited attention budget. Within this framework, "Purpose" is treated as a task description to be explicitly injected into the system prompt—fundamentally a supply-side information supplement, answering the question of "how to clearly tell AI what to do."

This assumption has merit, and has been widely validated in engineering practice—the long-horizon techniques covered by Context Engineering, such as compaction, structured note-taking, and sub-agent architectures, have genuinely helped large numbers of practitioners manage complex long-cycle projects. But across the 391 long-horizon collaborative sessions we documented, this path revealed a dimension it has not yet explicitly addressed: as the symbolic system became more precise, and historical discussion and current state continued to cohabit the same context, AI did not become more reliable—burdened by symbol parsing, it gradually abandoned genuine understanding of business semantics, retreated to self-referential reasoning within the symbolic layer, and generated outputs that appeared internally consistent but were physically disconnected from reality. We experienced this process in the first half of our own project, and have seen similar patterns being amplified in recent academic literature and new industrial tools. We name this failure **"Index Sickness."**

The emergence of Index Sickness points to a neglected dimension. Beyond the supply side, there is another side—**Consumer-side**: instead of trading quality by adding more constrained inputs to AI, trade quality by changing how AI activates information in the first place. All of this paper's work takes place on this side. The core operation on this side is precisely what is absent from the Context Engineering framework: rather than trying to feed AI to saturation, convey a genuine problem pressure—describe "what I am facing, what I need to solve" in natural language, and let AI translate this into concrete action intent on its own. Under the Transformer's semantic activation mechanism, such natural-language problem statements converge AI's interpretation of the semantic space, enabling it to activate purpose-relevant methodological knowledge without a retrieval intermediary. The supply side strives to "specify the purpose precisely"; the consumer side strives to "describe the problem clearly, and let AI understand the purpose itself"—the difference lies not in the surface action but in the underlying assumption: the former treats AI as an executor that needs to be fed; the latter treats AI as an understander that needs to be activated.

## 1.2 Research Method and Contributions

This study employs **Action Research (AR)** methodology: the researcher is simultaneously the practitioner, generating theory through iterative action in a real software development process. The paper's narrative structure corresponds to the standard AR cycle: Diagnosis ( § 2) → Theory Formulation ( § 3) → Action Design and Implementation ( § 4) → Evaluation ( § 4.3) → Reflective Learning ( § 5).

The research context is the "Bang-v3" project: a WeChat mini-program + H5 task-forwarding platform, spanning approximately one month and 391 collaborative sessions. The project was executed by one Owner collaborating with two AI roles (Chief-of-Staff CoS, Developer Dev), delivering a codebase and plugin system refactoring of approximately 80,000 lines of code across 506 files, while simultaneously building, iterating through multiple versions of, and writing summary documentation for the CSF (Collaboration Specification Framework)—approximately 900 files, 150,000 lines, including approximately 150 fully preserved raw session records.

Throughout the entire process, the Owner was exclusively responsible for "discussing business and software engineering" —no line of code or collaboration specification was written by hand. Each session began with the AI reading `context.md` to reconstruct its own working context—where the project came from, where it now stands, and where it is going; at session end, the AI updated `context.md` accordingly. No re-briefing was required.

Project session records and CSF specifications are partially open-sourced: 1. Code repository: https://github.com/huidev2025/CSF 2. Case documentation: https://huidev2025.github.io/CSF/cases/bang-v3/README_en.html

The paper's main contributions include:

1. **Naming and dissecting "Index Sickness"**: documenting the failure of supply-side formalization strategies in long-horizon LLM collaboration, naming the phenomenon, and analyzing its cognitive mechanism.
2. **Naming and recording the "Pang Principle (Semantic Vitality Law)"**: making an explicit statement and empirical test of the overlooked common truth that "natural language carrying explicit purpose conveys far greater information quality than symbolic expression" in the context of human–AI engineering collaboration.
3. **Proposing the consumer-side activation perspective**: identifying that activating AI's working state with natural language plus purpose constitutes a consumer-side strategy—converging AI's semantic interpretation space through explicit purpose, enabling the AI to preferentially activate purpose-relevant methodological knowledge and directly locate purpose-relevant project resources, without requiring a retrieval layer as intermediary filter. This constitutes a directional complement to the currently supply-side-dominated improvement path.
4. **Designing and validating "Baseline-Log Physical Separation"**: providing a lightweight, model-upgrade-independent cross-session state

management mechanism and reporting its effects in a real project.

### 1.3 Product and Methodological Declaration

The writing process of this paper constitutes a second case for the paper's argument—but what it demonstrates is not "Index Sickness was cured again" (the writing project was small in scale, never developed the illness, and lacks the conditions for comparison), but rather a real-time demonstration of "AI vitality" itself.

The collaboration mode throughout the writing process was as follows: the Owner was responsible for providing direction, rendering judgment, and raising corrections; the AI collaborator was responsible for maintaining complete cross-session context, participating in the reasoning chain, locating relevant content, and organizing written expression. Each of the Owner's interventions began with stating purpose in natural language—which is exactly the working state the Pang Principle describes. This paper is not only a paper about CSF, but also a document produced during CSF's operation; the two mutually corroborate each other.

It is necessary to state: all text in this paper was produced by the AI collaborator (GitHub Copilot, primary model version: Claude Sonnet 4.6, with Claude Opus 4.7 used supplementarily) under the above collaboration mode; the Owner did not independently write any paragraph. This fact must be disclosed as methodological information—not used as evidence for the argument. It is a transparency requirement, not an experimental design.

The complete session records, specification files, and the writing process of this paper have been partially open-sourced at https://github.com/huidev2025/CSF for independent verification. It should be emphasized that all empirical data reported in this paper—including session sequence numbers, file change records, Owner verbatim quotations, and systematic statistical analysis (classification statistics for 1,242 Owner inputs) —are supported by session records and session logs as evidence, and do not rely on AI's own statements as proof; the AI's role in the writing process is to organize expression from established context, not to generate empirical content.

---

# II. "The Bankruptcy of Formal Indexing": An Anatomy of Index Sickness

This chapter demonstrates, through representative failures in the Bang-v3 project's evolution, how AI develops cognitive bias in the absence of physical isolation and under conditions of excessive symbolization. The failure patterns can be organized along two dimensions: **spatial**—symbolic systems introducing cognitive burden and spurious closure within the current context; **temporal**—unpartitioned historical content penetrating across sessions to contaminate new tasks' semantic spaces.

In the project's early stages, the Owner had already observed AI's tendency to designate precise symbols and numbering systems for tasks, modules, and rules in order to build efficient, precise references. The Owner considered this something AI could do that humans could not, viewed it as effective, and tolerated the behavior. By around the 222nd session, the project had completed architectural design and entered intensive development; the symbol system had expanded rapidly, and the Owner began frequently asking "What does this symbol mean?", while the AI Chief-of-Staff's Instructions ballooned to **308 lines**, saturated with defensive symbolic constraints and incident-patch rules. Problems erupted collectively at the first integration test, and the Owner resolved decisively to upgrade CSF to shift to a predominantly natural-language indexing approach.

## 2.1 Symbol Proliferation and the Phantom Legislation Crisis

### Failure Phenomenon

At **Session 215**, this symbolic proliferation reached a concentrated breaking point. In that session, the AI Chief-of-Staff fully recited and "passed" a "Pending-Flow v2 Legislation Proposal"—the logic appearing complete and self-consistent at the symbolic level. However, when the Owner ran `grep` on physical files to verify, it emerged that AI had never established the corresponding physical skeleton in `CONTEXT.md`, and the developer-side documentation was also out of sync. AI had constructed a complete "completed state" in the symbolic layer while being entirely disconnected at the physical execution level. The Owner recorded a direct qualitative assessment:

> "Feedback is saturated with jargon words (such as SEC-2.0, D-139), which are noise to the Owner. The Chief-of-Staff used indexing and pre-writing mechanisms to reduce retrieval pressure, but in doing so evaded the core responsibility of 'understanding purpose.' Each time

> we hit a problem, we added a rule, causing Instructions to swell to 308 lines. These defensive patches not only occupy precious context space —they actually train the AI to fixate on indices, dissolving its capacity for independent thinking."

**Failure Mechanism Analysis**

The cognitive interference of symbolic systems on LLMs can be summarized in the following points:

1. **Multi-hop parsing introduces cognitive burden**: each time AI processes an identifier, it first needs to look up and map its meaning, then incorporate that into the current task. Repeated cross-file table lookups consume cognitive bandwidth that should be devoted to understanding business logic.
2. **Propagation of rigid errors**: once a symbol's mapping relation is incorrectly resolved, all subsequent reasoning based on that symbol will "correct" itself internally within the symbolic system, forming a fictional closed loop entirely detached from real business logic.
3. **Human correction pathways severed**: the high cognitive threshold of the symbolic system prevents the Owner—who holds business intuition—from judging by intuition whether AI's output is correct; errors are only discovered at physical execution.
4. **Literal Stickiness of historical context**: even when the Owner explicitly declares an old scheme deprecated, in an unreset long-horizon context, deprecated field names and interface structures maintain high weights in the Transformer self-attention mechanism's weight allocation, quietly infiltrating new design documents and creating cross-contamination between old and new terminology. For the self-attention mechanism, historical content in the context does not decrease in attention weight merely because it has been "declared deprecated"—physically clearing historical content is the only reliable means of blocking this contamination. Declaring "please ignore the old scheme" in the System Prompt is mechanistically ineffective. Engineering log D-108 from the project's 136th session records a canonical case of this mechanism: when the AI was wrapping up across sessions, it used residual impressions to complete the precise field label TD-01 as "024 Reverse Challenge Plugin," while the actual TD-01 = H5 Shared Component Library. This error spread to 13 cross-file labels; the session was classified as a "meta-incident repair session," with all development progress halted and dedicated to investigation and correction.

This phenomenon has received quantitative support from multiple independent experimental studies. Chroma's systematic tests across 18 mainstream LLMs demonstrate that even under conditions of ideal retrieval recall rates, distractors in the context still cause non-monotonic degradation in task performance [7]; Raju et al. further found that after forcibly extending context, even when all relevant content has been injected, agent task success rates still drop sharply [8]. These results indicate the problem lies not in retrieval precision but in LLMs' own context attention processing mechanisms.

In the subsequent Session 227, when AI again attempted to describe methodology using an abstract symbolic deductive system (`C1-C9 → P1-P4 → E1-E8`), the Owner directly critiqued this pattern:

> "It's still the SP5 symbol-sickness story. AI easily builds up a complex symbolic system of its own, then references it back and forth, gets carried away, and thinks that is mathematics. Mathematics is not theory; in my view, mathematics or symbols unrelated to business are useless."

In Session 228, the Owner further offered a diagnosis at the cognitive mechanism level:

> "In the SP5 phase, the Chief-of-Staff built many symbols and markers to help itself index. Once the indexing became complex and citations increased, it obviously couldn't keep up. Storage capacity was still sufficient, but it had to ensure retrieval efficiency, so it gave up on retrieval." [That is: to ensure retrieval efficiency, it chose to forgo some retrieval and original-text verification—editor's note]

The failures presented in the cases above take different forms but share a single underlying dynamic: the Transformer attention mechanism allocates weights by statistical co-occurrence patterns and cannot identify a symbol's "validity status"—once a symbol enters the context, it continuously participates in activation regardless of whether its business meaning remains valid. When symbolic systems form high-density patterns in the context, attentional resources concentrate in the symbolic layer and real business semantics are structurally marginalized—whether it is the disconnected reasoning of Phantom Legislation, the bandwidth consumption of multi-hop parsing, or the cross-session contamination of Literal Stickiness, all are projections of this dynamic in different scenarios.

These failure patterns are not accidental engineering incidents, nor bugs in specific LLM versions. It should be noted that we are not claiming symbolic systems are the sole cause of these failures—action research methodology does not require excluding alternative explanations; our claim is that within this project's action-reflection cycles, the shift from engineering intuition (symbols disambiguate) to engineering observation (symbols amplify ambiguity) itself constitutes an engineering finding worth explicit recording, collectively pointing to a structural limitation of the formalization path. The next chapter builds on these engineering observations to formulate a theoretical explanation for the common root cause of these failure phenomena.

---

# III. From Engineering Observation to Theoretical Proposition: The Formation of the Pang Principle

Running through all the failure cases above is a recurring baseline: the more symbols, the less AI understands. This observation points to a more fundamental question: **How exactly does an LLM process contextual information such that adding more precise symbols actually increases cognitive burden?**

## 3.1 Engineering Axiom: The Semantic Information Advantage of Natural Language

CSF's engineering design has always been grounded in a consciously held foundational intuition. Around Session 215, the Owner progressively articulated this intuition as an explicit theoretical proposition, which was explicitly recorded as the framework's Engineering Axiom:

> "Also, notes should be in natural language, because natural language is the medium that carries the maximum information, and can be fully utilized by you at each retrieval. This is the exact opposite of RAG." (Session 215)

In Session 229, this intuition was explicitly stated as a theoretical proposition:

> "Natural language expression carries far more information than symbolic systems. Because natural language expression can be interpreted in context—it is inherently adaptive. This is CSF's true first-principles question."

Engineering Axiom: Natural language expressions carry far greater information than symbolic systems.

(**Note**: The term "Engineering Axiom" here follows the action research tradition—it denotes a foundational proposition explicitly named by inductive practice, not a premise requiring no external proof within a deductive system. The independent experimental studies cited below are intended to seek external corroboration, not self-justification.)

The core argument is: symbols must first undergo a formal "look-up and mapping" conversion before they can be processed, truncating the model's native semantic activation path; natural language, by contrast, directly activates the model's high-dimensional prior knowledge, conveying more effective semantic information with fewer tokens.

## 3.2 Fact ①: The Tradeoff Between Intelligence and Determinism

Fact ①: The value of artificial intelligence comes from "intelligence"; attempting to make it reliable as machinery is eliminating its value.

This fact identifies an intrinsic contradiction in current industry practice: the very reason for adopting large language models is to harness their flexible semantic reasoning capability; yet the direction of pursuing determinism through formal constraints runs directly counter to this purpose. Mandating formatted output is a canonical manifestation of this contradiction: AI is forced to divide limited attention between "format compliance" and "semantic accuracy"—the more precise the format constraint, the more it crowds out semantic comprehension. At the extreme, AI generates output that is formally perfectly correct and substantively completely wrong—which is precisely a variant of the Phantom Legislation phenomenon.

## 3.3 Fact ②: The Semantic Dilution Law

Fact ②: Over time, the information carried by an expression will inevitably grow, and its semantics will be diluted.

The Owner elaborated on this law in Session 228 with the "cloth doll analogy":

> "As long as information volume increases, conceptual load will be diluted. A single concept's range of interpretable meanings expands, but the intensity of reference weakens. Simply put, in a novel, a noun cannot maintain one consistent meaning throughout... Just like a cloth doll: the longer it has been with me, the more stories it accumulates, the more varied the experiences and feelings become. This is unavoidable."

This analogy precisely captures the internal structure of the Transformer self-attention mechanism: the historical contexts accumulated by a concept do not fade from weight allocation because a human declares it "deprecated"—this is exactly the root source of **Literal Stickiness**. The longer a concept has been referenced, elaborated, and discussed, the more diffuse its semantic activation boundary in the attention matrix, and the more the intensity of its original referent is diluted.

This observation directly challenges the "larger context window solves everything" assumption: physically expanding the window does not prevent conceptual semantic dilution, and may even accelerate the process by accommodating more historical discussion. Therefore, **actively selecting and maintaining a bounded semantic space around concepts—rather than unrestrainedly expanding windows—is the correct design orientation for preserving semantic precision.** Liu et al.'s systematic experiments on the "Lost-in-the-Middle" phenomenon in long contexts [9] demonstrate that LLMs have systematic attentional amnesia for information in the middle of contexts; Chroma's large-scale cross-model experiments [7] reveal that this degradation is unavoidable even on the simplest tasks. These experimental results quantitatively support the necessity judgment for semantic space control.

### 3.4 Purpose as the Control Variable

The Engineering Axiom establishes natural language's information advantage, but natural language's interpretive space is open. Without constraint, vitality also produces ambiguity and deviation. Therefore, a minimal control variable is needed to converge AI's semantic understanding to the boundaries of the current task. We establish "purpose" in this role through the following reasoning chain:

1. Natural language carries maximum information (Axiom).
2. But an open interpretive space produces ambiguity (Deduction I).
3. A minimal control variable must be introduced to converge interpretive direction without sacrificing expressive elasticity (Deduction II).
4. "Purpose" semantically corresponds to the core constraint of the current task and is the most efficient convergence control variable (Deduction III).

This yields the semantic space control formula:

$$\text{Natural language} + \text{Purpose} = \text{Maximum information quality}$$

$$\text{Natural language} - \text{Purpose} = \text{Ambiguity noise}$$

As the Owner articulated in subsequent theoretical synthesis:

> "Natural language + purpose not only activates information utilization efficiency from the consumer side; more importantly, it activates AI's vitality with the correct information."

The supply-side/consumer-side distinction established earlier is here developed at the theoretical level. **Supply-side** strategies share the common assumption of "the more complete and determinate the input, the more reliable the output"—symbolic identifier systems, rule stacking, RAG injection, and expanding context windows all belong here. RAG is also supply-side—although its core mechanism is selecting which documents to inject into AI, the AI's cognitive activation still comes from externally injected document fragments rather than its own prior knowledge directly activated by purpose; "retrieval layer intervention" is itself the characteristic of a supply-side operation. **Consumer-side** strategies move in the opposite direction: using explicit purpose to converge AI's semantic interpretation space, enabling AI to preferentially activate purpose-relevant methods within its own knowledge probability space and directly locate purpose-relevant project resources—without retrieval layer intervention, without secondary filtering, because AI knows the purpose and naturally knows what to read and what not to read. The engineering significance of the Engineering Axiom and two Facts stated above lies precisely here: they are not a filter on the supply side, but an activator on the consumer side.

We name the Axiom together with its two deductions (Fact ① and ②) and the introduction of the purpose control variable as the **Pang Principle (Semantic Vitality Law)**. The character "乓" (Pāng) is taken from the Chinese colloquial image "pāng de yīshēng"—the instant a self-evident truth thuds into place; the name is a deliberate reminder to readers: the content of this principle is not unfamiliar; what truly warrants examination is the reality of its systematic neglect. Beyond the misunderstanding that it "violates Shannon information theory," many readers will find this principle unremarkable—all common sense. Yes: this paper's empirical record is precisely a reminder that in the context of LLM collaboration, research efforts and industrialization paths that violate these common truths are being systematically advanced in a highly organized, self-reinforcing manner—each advance adding another layer of shackles on "AI vitality."

The above observations and propositions together point toward a question to be tested: oriented toward the consumer side, can an operable engineering mechanism be found at the implementation level—and does this mechanism actually work? The next chapter documents precisely this verification.

It is worth noting in advance: this mechanism turns out to be remarkably simple—introducing no new middleware, only adjusting the physical organization structure of files. We believe this simplicity is itself a notable fact. When the design direction aligns with the system's internal operational logic, implementation tends not to consume large amounts of friction; conversely, continuously growing engineering complexity can sometimes be precisely the signal that the direction contains an error. This is not proof of correctness, but in our view can serve as a corroborating side evidence for directional judgment.

---

# IV. Physical Defense Mechanism: Baseline-Log Physical Separation

## 4.1 Design Principle

From the theory above, we establish the core principle of mechanism design: **do not rely on Instructions within the Transformer window to resist noise; instead, physically block the noise's input source at the file organization layer.**

This mechanism introduces no middleware code, requires no model upgrade, builds no graph network—it only adjusts the physical organization structure of files. This makes it a purely **consumer-side design**: it adds no new constrained input to AI, only manages the information boundary AI activates when each session starts.

The establishment of this principle came from direct practical recognition. The Owner identified in Session 215:

> "In the extended cognition system, decomposition is your core capability—grounded in understanding purpose, and actively chosen. The extended cognition system is nothing more than a task to be solved. That task's accurate description: for a specific purpose, build a specific information indexing system, and prepare to minimize context space requirements during task execution."

The project specifications also record a counterexample from before this principle was established: in the early stage when DEVNOTES simultaneously carried both "settled decisions" and "process discussion brainstorming," when AI reloaded the file in subsequent sessions, it could not distinguish between "conclusions" and "rejected discussion options," and incorporated deprecated schemes into new design outputs. The project engineering specification document (bang-v3/plan-csf-v2/protocols/closing-discipline.md § 6) explicitly records this as a counterexample: "DEVNOTES simultaneously carrying 'decision summary' and 'process discussion log' → cognitive-layer baseline contaminated by process information. Correction: decision summary retains only conclusions; discussion moved to execution-layer log." This direct observation catalyzed the core writing specification at the heart of Baseline-Log Separation.

## 4.2 Mechanism Design

Baseline-Log Separation serves two parallel purposes, neither dispensable.

First, **ensure working agents contact only current concepts, uncontaminated by history**—the reason the baseline exists. The baseline uses overwrite-mode updates, retaining only current truth at any moment, carrying no trace of historical discussion.

Second, **decision process continuity must not be broken**—the reason session logs exist. Subsequent workers need to know how a decision was made, what options were considered, why some were eliminated. Without process, every historical decision becomes a black box—whenever a new situation requires backtracking, one cannot understand why something was done, nor evaluate whether it can be adjusted.

As the Owner put it: **"Move the current state forward, and a log naturally appears."** Baseline and session log are not two independently maintained systems—they are slices of the same information body across two temporal dimensions: the baseline is the "cross-section of now"; the session log is "the path walked."

**Baseline (e.g., `context.md`)** — **Overwrite-mode updates.** The baseline stores only the current task triple: **Purpose + Method + Resources**. Purpose comes first—the AI first understands purpose, then selects methods based on its own knowledge, then organizes just-sufficient resources. After each task, the baseline is overwritten with the new state, retaining no history. This locks new-session cold-start input at constant scale; startup cost does not grow regardless of how many sessions the project has undergone.

**Session Log (e.g., `session-NNN.md`)** — **Append-mode recording.** The session log's purpose is **reconstructing the context of each decision and action.** It records each session's process in append mode: how decisions formed, alternatives considered and eliminated (with rationale), resource references (including filenames, folder names, section names, and other symbolic indices), and business-understanding corrections arising from the work. Natural-language writing is a hard rule—not excluding the appearance of symbolic indices like filenames, but requiring every entry to have natural-language explanatory support, enabling subsequent workers to reconstruct context through reading comprehension rather than mechanical retrieval.

Natural-language writing is the key. When backtracking is needed, what AI reads is not keyword indices but the complete decision context of that moment—why this was done, what options existed, why one path was chosen. AI judges from contextual understanding, not keyword matching; when understanding is correct, conclusions are naturally accurate. This is the essential mechanism by which session logs function—and is the temporal extension of natural language's advantage as an information carrier over symbolic systems.

Session logs do not physically enter new-task context windows, blocking interference from deprecated content on new tasks.

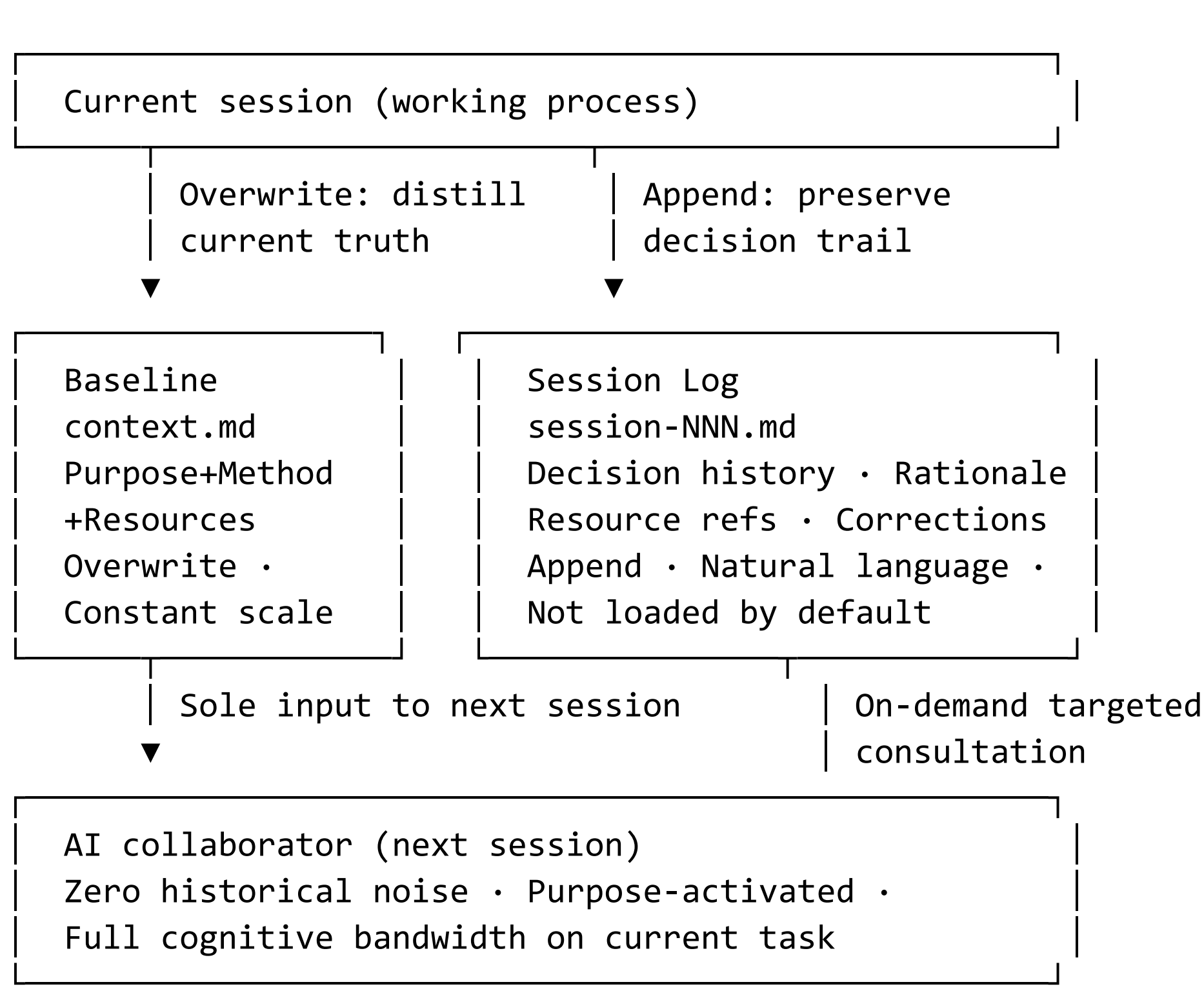

## 4.3 Results

By moving historical incident-response patches and symbolic index systems out of the context, post-Session 215 AI Instructions dropped from **308 lines to ~80 lines** (~75% reduction), achieving "thin-shell" status. This independently corroborates SkillReducer [10], which validated a similar principle at the agent tool-description layer: iteratively pruning the tool candidate set and simplifying descriptions significantly improves task execution accuracy—danger comes from redundant information in the window, not insufficient information.

It is worth noting that this project went through **two rounds** of specification simplification. The first round ("SP-Trim," ~Sessions 136–141) applied a quantitative reduction strategy, cutting specification lines from ~1,100 to

~480 (↓56%), yet the core context file (`context.md`) actually grew from 2,197 to 2,256 lines—and Index Sickness did not disappear. The genuine cliff-drop occurred in the second round: the Baseline-Log Physical Separation described in this paper. This contrast reveals that the effective variable is not the line count of specifications, but the organizational architecture of information. Full discussion in §5.1.

Key comparisons before and after upgrade:

| Dimension | Before (symbol indexing + rule containment) | After (baseline isolation + natural language) |
|---|---|---|
| Instructions scale | 308 lines (heavy defensive patches) | ~80 lines (meta-capability pointers + collaboration discipline) |
| DEVNOTES decision entries | 141+ entries (with B/A/M three-tier classification and reverse index) | 5 entries (one-line natural language description, no classification system) |
| Owner correction density | 0.79 corrections/session (#136–#214, 57 session files) | 0.53 corrections/session (#253–#395, 38 session files), ↓33% |
| Phantom Legislation | Recurrent (Session 215 representative) | No instances matching Index Sickness failure patterns were observed in the subsequent ~150 session logs |
| Human correction capability | Symbol-dependent; intuition cannot intervene | Natural language; Owner can verify by intuition at any time |

> **Note**: This project did not systematically record per-session token consumption; context windows were generally 160K–200K tokens, and sessions typically ended before window exhaustion, so quantitative context-load comparison cannot be extracted from the logs. However, the session logs and session records preserve an observable qualitative indicator: the operational meaning of "no instances observed" here is that across the subsequent ~150 sessions, the Owner never encountered any session in which "AI claimed completion but physical file verification found it disconnected," nor did any meta-incident

> repair session arise that needed to be dedicated entirely to error correction (of the type of Session 215 in §2.1). Concretely: after the upgrade, the Owner almost never needed to stop mid-session to point out errors or restate collaboration discipline—the interaction pattern simplified to "start, confirm, end." Before the upgrade, frequent correction, clarification, and realignment were session norms. This shift is traceable in the session logs and session records and can be treated as a qualitative indicator of substantially reduced collaboration friction. This limitation is a direct consequence of the action research design—data collection served engineering rather than experimental purposes.

The Owner correction density figure in the table above is derived from a systematic statistical analysis of 139 session record files covering sessions #136–#395. We extracted all Owner inputs (`User:` lines), totaling 1,242 entries, and applied a keyword-matching plus manual review classification method to identify "error correction" and "business course-correction" inputs, with suspected Index Sickness cases individually verified against the operational definition in §2.1. Classification rules, manual review records, and raw statistical data have been open-sourced alongside the project for independent reproduction (same repository as §1.2; analysis scripts located under `cases/bang-v3/`).

## 4.4 Supplementary Case: Cross-Session Provenance Efficiency (Sessions 373–375)

The mechanism's value received further validation in a systematic gap analysis during the project's later stage.

**Background.** At integration testing in Session 373, the team found the "Forwarding Overview Bar" component had never been implemented. On the surface this was a single-point feature gap, but deeper investigation revealed a root cause spanning four design layers: two related design theme packages (TD-02 and Main-5) had advanced independently at different times, and when one was finalized, the other's business concepts did not yet exist—interfaces never aligned.

**Provenance.** To locate this temporal disconnect, the AI CoS needed only to read two files based on the relevant task's resource index: TD-02's `package.md` and Main-5's planning file. Both files, per CSF convention, recorded creation dates and session numbers at inception. From these two semantic annotations, the AI gave an immediate precise verdict:

> "TD-02 created: 2026-05-19 (Session 138); Main-5 'Forwarding Relationship Rendering' established: 2026-05-21 (Session 168). When TD-02 was finalized, Main-5 simply did not exist."

This judgment spanned 30 sessions of temporal distance—without scanning 370+ raw session logs, without vector retrieval (RAG), completed within a single session.

**Mechanism insight.** This case illustrates another dimension of the session log mechanism: physical isolation's purpose is not only "blocking noise" but equally "preserving context." Session logs record not a chronological stream but the complete decision context of each action—including purpose, alternatives considered, and rationale for choices—all in natural language, enabling AI when backtracking to reconstruct decision context and judge from understanding rather than keyword matching.

This is the consistent logic of the entire mechanism: not pursuing mechanical precision in AI, but enabling full understanding—when understanding is correct, conclusions are naturally accurate. In Session 373's case, AI read two files carrying complete decision context to complete temporal localization spanning 30 sessions—not through retrieval technique, but through understanding.

This is consistent with the baseline's design philosophy: **higher semantic information density is superior to accumulated information volume.**

Looking upward from the tool layer, the core proposition this mechanism reveals is: what determines long-horizon human–AI collaboration quality is not AI memory capacity, but the semantic control structure jointly maintained by the human–AI team. The theoretical implications of this proposition and the larger research landscape are the central topic of §5.3.

---

# V. Discussion and Conclusion

## 5.1 Alternative Explanations and Validity Argument

The most prominent alternative explanation for this study is the **team learning effect**: after experiencing intensive failures in the first half of the project, the Owner and AI collaborator may have naturally become more cautious in their collaboration patterns—rather than the Baseline-Log mechanism itself bringing improvement. This paper's observations cannot strictly exclude this effect, because "mechanism introduction" and "experience accumulation" coincide temporally. We present four traceable counter-arguments.

**Counter-argument 1: The structural fact of two rounds of simplification.** As described in §4.3, the project had already undergone one round of quantitative reduction ("SP-Trim," ~Sessions 136–141) before the CSF upgrade. The results of both rounds are compared below:

| | **Round 1: SP-Trim (~Sessions 136–141)** | **Round 2: Baseline-Log Physical Separation (~Sessions 215–252)** |
|---|---|---|
| Core strategy | Quantitative reduction (delete rules, merge files, cut lines) | Change information architecture (baseline overwrite / log append) |
| Specification line change | ~1,100→480 lines (↓56%, with quantitative acceptance record) | Instructions 308→80 lines (↓75%) |
| `context.md` lines | 2,197→**2,256 (+2.7%, actually grew)** | 680→**189 (↓72%, cliff-drop)** |
| DEVNOTES decision entries | 141+ entries (unchanged) | 141+ entries→**5 entries (↓96%)** |
| Post-upgrade Index Sickness | **Still erupted** (#204 severe error, #205 restart, #215 Phantom Legislation, #221 failure) | **0 cases** (none observed across subsequent ~150 sessions) |

At that point, the team had already accumulated ~140 sessions of collaboration experience, and the specification volume had already been cut by more than half. If improvement came from "experience accumulation," it should have been observable after Round 1—but in fact, Index Sickness erupted in a concentrated burst across the ~60–80 sessions following Round 1, and only disappeared after Round 2 changed the information organization structure. The improvement corresponds to a qualitative change in information architecture, not to accumulated experience or reduced rule count.

**Counter-argument 2: Quantitative shift in correction density.** As reported in §4.3, systematic statistics over 1,242 Owner inputs show correction density dropped from 0.79 corrections/session pre-upgrade to 0.53 post-upgrade (↓33%); of the 45 genuine corrections pre-upgrade, 9 (20%) met the Index Sickness operational definition; of the 20 post-upgrade, 0 did. The learning effect may explain a natural decline in total corrections, but it cannot explain "cases meeting the Index Sickness operational definition dropping from 9 to zero"—the latter corresponds to a structural disappearance of a specific failure mode, not a gradual improvement in overall quality.

**Counter-argument 3: Qualitative shift in session titles.** Pre-upgrade session logs (before Session 215) frequently bear titles such as "Failure," "Severe Error Discovered," "Start Over—Redefine Collaboration Norms," and "Talked Into a Dead End"; post-upgrade titles are uniformly simple and productive ("Development"), with none recording correction or reconstruction episodes.

**Counter-argument 4: Qualitative shift in Owner input content.** Pre-upgrade, the Owner frequently issued multi-point corrective instructions (canonical form: "I have three issues: 1. Engineering-level question⋯ 2. You need to fully understand [the theory] before proceeding—don't just mechanically execute⋯. 3. Analyze the problem in the business context"), and repeatedly interrupted sessions to address "severe errors"; post-upgrade, Owner inputs contracted to brief directional prompts ("Let's start," "Continue"), with no diagnosis or correction content required. If this improvement were due to accumulated prompting skill, we would expect prompts to become more *refined* but remain the same *type* (correction still present, just more precise); what was actually observed is that the *necessity* of corrective input disappeared altogether. This qualitative shift differs mechanistically from the learning effect's expected pattern of "handling the same type of problem more skillfully," and is traceable in the session records. The improvement was not gradual—also inconsistent with the gradual nature of learning effects.

Of the four counter-arguments, the first two are derived from quantifiable project records and the latter two from qualitative log observations; all four are independent of each other, and all point to the same conclusion: the improvement corresponds to a structural change in information architecture, not a natural accumulation of collaboration experience. That said, multi-project comparison studies remain the ultimate path to statistically distinguishing the two effects.

## 5.2 Research Limitations

### Methodological Limitations

This study uses single-case action research, with data from one project (Bang-v3) over a specific time period. Action research has recognized research value in software engineering [11][12], and Treude and Storey note that in software engineering research involving deep LLM participation—where "AI is not only a tool but an active collaborator"—the researcher-as-practitioner action research framework captures cognitive phenomena in real interaction better than traditional controlled experiments [13]. The generalizability of conclusions requires validation across more projects, more collaborators, and more LLM versions. Some effect indicators come from participant observation rather than controlled quantitative measurement; subsequent research can design more rigorous evaluation protocols.

**Reflexivity Note.** The first author is simultaneously the project Owner, the designer of the CSF mechanism, and the evaluator of the outcomes reported in this paper. This triple role carries the risk of confirmation bias—the Owner had a motive to see the mechanism they designed validated by the project they ran. The bias-reduction measures adopted include: (1) session records from the 136 sessions onward are fully preserved for independent third-party verification, rather than selectively presented; (2) the evaluation baseline is not subjective satisfaction but traceable engineering facts—Instructions line count is measurable, correction events within sessions are locatable, and Phantom Legislation cases include `grep` command records of physical file verification; (3) the Baseline-Log Separation mechanism itself requires the baseline to contain only conclusions and session logs to contain the full process—this physical constraint also limits the Owner's ability to revise the narrative post hoc. Despite these measures, confirmation bias cannot be entirely eliminated; subsequent multi-project, multi-Owner independent validation is necessary.

**Theoretical Scope**

The axiom that "natural language outperforms symbols" holds within the context of LLMs pretrained primarily on natural language. For specialized models pretrained primarily on code or structured data, the scope of this axiom requires separate evaluation.

## 5.3 Conclusion

This paper's most important finding is not that "symbol systems sometimes fail"—engineering practitioners have long had this intuition. What truly deserves examination is the **universality and unconscious nature** of this failure: in the first half of our project (before the upgrade around Session 222) and in the academic literature, new industrial methods, and tools that have emerged in recent years, the formal-constraint path is being advanced in a highly self-reinforcing manner—every time frustration arises over AI's unreliable behavior, the instinctive response is "add another rule," "refine a symbol further"; most related papers and industrial tools propose improvements centered on the same core question: how to optimize the information injected into the context window. No one is deliberately making a mistake, but as a whole, the field is broadly departing from an overlooked common truth: **the primary obstacle to maintaining semantic accuracy is historical noise in the active context that should not be there—not insufficient AI memory.**

This paper's contribution is not discovering this common truth—it was always there. The contribution is: **re-raising and naming it (the Pang Principle), documenting the consequences of violating it (Index Sickness, with complete session records), and demonstrating the effects of returning to it (before vs. after physical isolation: Index Sickness cases from 9 to zero, correction density down 33%)**. These three elements constitute a complete consumer-side argument chain: the common truth first, the cost of violating it second, the way back third. Against the backdrop of the supply-side improvement path continuing to escalate, this chain points to a directional alternative—not a negation of existing work, but a complement to a neglected dimension. In recent years, the heavy structured-memory architectures represented by CoMem, REAL, and GAM [3][4][5], and formalization attempts such as Git Context Controller [2], collectively confirm that long-horizon context degradation is a genuine engineering challenge; but their prescriptions—more complex middleware, more precise symbolic structures—run counter to this paper's direction. They operate under the assumption of "how to help AI remember more." This paper's diagnosis is the inverse: the problem is not that AI remembers too little, but that the active context is saturated with historical noise that should not exist—the divergence in their prescriptions is rooted in the formulation of the problem. Whether the two directions are complementary rather than opposed is not yet answerable from available experimental data—this is among the most immediate topics for follow-up work. The industrial practice direction represented by Context Engineering [6] addresses a supply-side question: how to dynamically select what information to inject into and remove from the context window—including compaction, externalized note-taking, and sub-agent architectures. This is adjacent but non-overlapping with the structural problem identified in this paper (the physical cohabitation of baseline and historical discussion); the two exist as complementary dimensions forming the complete engineering landscape of long-horizon human–AI collaboration.

This finding has a clear home in the larger intellectual landscape. Clark and Chalmers' extended mind theory [14] and Hutchins' distributed cognition research [15] jointly indicate that cognitive processes do not terminate at the boundary of the individual mind but extend into the system constituted by artifacts, documents, and the surrounding environment. In the human–AI collaboration context, baseline and session logs are the physical carriers of this extended cognitive system—not patches on AI memory, but the shared cognitive control structure of the entire human–AI team. Booch identified, before the LLM era, that software development is fundamentally a social activity: establishing shared mental models across teams is more fundamental than any formal specification [16]. When one member of the collaboration team is a large language model, the weight of that judgment only increases. The engineering findings documented in this paper represent a concrete engineering instantiation of this philosophical position in the human–AI collaboration context—with complete session logs and session records available for external verification.

This study is built on single-project action research data; its generality awaits validation across more projects, collaborators, and LLM versions. The effectiveness boundary between the "active selective forgetting" strategy and heavy memory architectures lacks controlled experimental data—this is the most direct next step. The deeper question: as semantic control structure becomes a core asset of human–AI collaboration engineering, how will its design principles evolve alongside continuous model improvement? What the Pang Principle reveals is not merely a set of specific engineering practices, but a class of problem formulations—a class that becomes more, not less, serious as models grow more capable.

---

## Acknowledgments

The authors warmly thank Shuren Song for his companionship and countless weekend exchanges of ideas; his theoretical guidance and emotional support were a core driving force behind this work.

This paper was produced under the CSF (Collaboration Specification Framework) Baseline-Log Physical Separation protocol. Throughout the writing process, the human author (H. Zhang) was responsible for purpose activation—stating direction verbally, rendering judgment, and raising corrections; all manuscript prose was generated by the AI collaborator. The authors declare this AI involvement in full, consistent with the methodological declaration in § 1.3.

---

## References

### Formalization Constraint Path

**Heavy Structured-Memory Path**

**Context Engineering (Framework Overview)**

**Empirical Studies on Long-Context Limitations**

**Empirical Support for the "Subtraction Strategy"**

**Action Research Methodology**

**Extended Cognition / Cognitive Science / Socialized Software Engineering**

# 这篇文章由AI写就：在连续391次会话中由AI管理语义空间并消除"索引病"

**张慧(Hui Zhang)**
深圳市云溪科技有限公司
zhanghui@ecloudriver.com

**宋树仁(Shuren Song)**
清华大学信息化技术中心
songsr@tsinghua.edu.cn

## 摘要

在应对大语言模型(LLM)长程协作中的概念漂移问题时，当前业界普遍的工程直觉是：用更精密的形式化约束来换取更可靠的输出——为任务实体设计符号代号系统，在 System Prompt 中不断堆叠防错规则，以更大的上下文窗口容纳更完备的约束索引。本文的工程记录表明，这一方向在长程场景下，实际效果可能与设计预期相反。

本文在一个历时约一个月、391 次协同会话的真实软件重构项目(帮找 v3)中，采用行动研究方法，记录并分析了上述策略在本项目中的失效过程。当符号体系超过一定复杂度后，LLM 并不因约束更精密而变得更准确——它会放弃对真实业务语义的理解，转而在符号层进行自我循环论证，生成表面自洽、物理挂空的虚假输出。我们将这一失效模式命名为 **"索引病(Index Sickness)"**，其典型表现称为 **"虚立法"**。

这一失效现象有一个反直觉的解读：长程 LLM 协作的核心障碍，不是 AI 记忆的不足，而是活跃上下文中存在了不该存在的历史噪声。认识到这一点，解法变得简单——它本质上只是对一条被忽视的常识的回归。本文将这一常识命名为 **"语义活性定律(乓定律，Pang Principle)"**：在人与AI协作中，携带明确目的的自然语言，其信息质量远高于符号表达。基于这一认知，本文设计并验证了其物理落地机制：**"基线-Log 物理隔离(Baseline-Log Separation)"**。在同一项目中，该机制将 AI Instructions 规模缩减约 75%，在后续约 150 次会话中，未再观察到索引病现象；Owner 纠偏密度下降 33%，升级前确认的 9 例索引病案例在升级后归零。

---

# 一、引言

## 1.1 问题背景与本文视角

近年来，AI 辅助软件工程领域涌现出一批共同指向同一方向的改进：更精密的符号标识符系统与 Prompt 工程规范化 [1][2]、更重型的结构化记忆架构 [3][4][5]、更大的上下文窗口与检索增强注入。这些工作回答的是同一个问题——**如何向 AI 提供更好的输入**。我们将这一类路径统称为**供给侧(Supply-side)** 改进：其共同假设是，AI 的输出质量由供给侧的输入完备性和确定性决定，输入越精密越完整，输出越可靠。

这些改进路径近年来被工业界整合为统一的实践方向，Anthropic 将其命名为**"上下文工程(Context Engineering)"** [6]："从持续演进的信息全集中，为每次推理选取最小高信号 Token 集，以最大化目标行为的概率"——覆盖系统提示设计、检索增强注入、长程压缩与结构化记忆等路径，统一的诉求是管好模型有限的注意力预算。在这一框架下，"目的"被视为一条需要在系统提示中显式注入的任务描述，本质是供给侧的信息补充——回答的问题是"如何向 AI 说清楚要做什么"。

这个假设有其合理之处，Context Engineering 所覆盖的压缩、结构化笔记与子智能体协作等长程技术，也在大量真实项目中得到了工程验证。但在我们记录的 391 次长程协同会话中，这一路径展示了一个它尚未显式处理的维度：随着符号系统的精密化，当历史讨论与当前状态持续在同一上下文中混居时，AI 并没有变得更可靠——它在符号解析的负担下，逐渐放弃了对真实业务语义的理解，转而在符号层进行自我循环论证，生成表面自洽、物理挂空的虚假输出。我们在自己项目的前半程经历了这一过程，并在近年学术文献与工业界新工具中看到了相似的模式被持续放大。我们将这一失效命名为 **"索引病(Index Sickness)"**。

索引病的发生提示了一个被忽视的维度。供给侧之外，存在另一侧——**消费侧(Consumer-side)**：不通过向 AI 增加约束输入来换取质量，而是通过改变 AI 激活信息的方式本身来换取质量。本文的全部工作都发生在这一侧。这一侧的核心操作，正是 Context Engineering 框架中缺席的那件事：不试图喂饱 AI，而是传递一个真实的问题压力——用自然语言描述"我在面对什么、我要解决什么"，让 AI 自行将其翻译为具体的行动意图；这种陈述在 Transformer 的语义激活机制下能收敛 AI 对语义空间的解读，使其无需检索层中介即可激活目的相关的方法知识。供给侧在力图"把目的说精确"；消费侧在力图"把问题说清楚，让 AI 自己理解目的"——两者的区别不在表面动作，在底层假设：前者视 AI 为需要被喂饱的执行器，后者视 AI 为需要被激发的理解者。

## 1.2 研究方法与贡献

本研究采用**行动研究(Action Research)** 方法：研究者同时是实践者，在真实软件开发过程中，通过迭代行动产生理论。本文的叙述结构对应 AR 的标准循环：诊断(§2)→ 理论提炼(§3)→ 行动设计与实施(§4)→ 评估(§4.3)→ 学习反思(§5)。

研究背景为"帮找 v3"项目：一个微信小程序 + H5 任务转发平台，历时约一个月，完成 391 次协同会话。项目由一名 Owner 与两个 AI 角色(参谋长 CoS、开发者 Dev)协作完成，期间交付了约 8 万行代码、506 个文件的底座与插件系统重构，同时完成了 CSF(Collaboration Specification Framework，协作规范框架)规范的构建、多版本迭代与总结文档撰写(约 900 个文件、15 万行，含原始会话记录完整保存约 150 份)。

整个过程中，Owner只负责"聊业务、聊软件工程"，没有动手写一行代码或协作规范文档。每次会话开始，AI 阅读 `context.md` 自建工作语境——知道项目从哪儿来、现在在哪儿、准备到哪儿去；会话结束时，AI 自动更新 `context.md`。Owner 无需每次重新介绍项目背景。

项目部分会话记录与 CSF 规范已开源： 1. 代码仓库 https://github.com/huidev2025/CSF； 2. 案例文档 https://huidev2025.github.io/CSF/cases/bang-v3/README_en.html 。

本文的主要贡献包括：

1. **命名并剖析"索引病"**：记录供给侧形式化策略在长程 LLM 协作中的失效现象，为其命名，并分析其发生的认知机制。
2. **命名并记录"乓定律(语义活性定律)"**：在人机协同工程语境下，对"携带明确目的的自然语言，其信息质量远高于符号表达"这一被忽视的常识作出显式表述与实证检验。
3. **提出消费侧激活视角**：指出以自然语言加目的来激活 AI 的工作状态，属于消费侧策略——通过明确目的收敛 AI 的语义解读空间，使 AI 优先激活目的相关的方法知识，并直接定位目的相关的项目资料，无需经由检索层的二次筛选。这与当前以供给侧为主导的改进路径构成方向性补充。
4. **设计并验证"基线-Log 物理隔离"机制**：提供一种轻量级、不依赖模型升级的跨会话状态管理方案，并报告其在真实项目中的效果。

## 1.3 产物与方法论声明

本文的写作过程构成论文论点的第二个案例——但它证明的不是"索引病被再次治好"(本文写作规模较小，且从未发病，不具备对比条件)，而是对"AI 活性"本身的一次实时演示。

整个写作过程的协作模式如下：Owner 负责提供方向、作出判断、提出纠偏；AI 协作者负责维护跨会话的完整语境、参与论证链的推演、定位相关内容、组织文字表达。Owner 的每一次介入，都以自然语言表述目的为起点——这正是乓定律所描述机制的工作状态。本文不仅是关于 CSF 的论文，也是在 CSF 运行过程中产出的文档，两者互为印证。

此处需要说明的是：本文所有文字均由 AI 协作者(GitHub Copilot，主要模型版本：Claude Sonnet 4.6，兼用 Claude Opus 4.7)在上述协作模式下完成，Owner 未独立撰写任何段落。这一事实需要作为方法论信息明确披露，而非作为论点的证

据使用——它是透明度要求，不是实验设计。

完整的会话记录、规范文件与本论文的撰写过程，已部分开源于 https://github.com/huidev2025/CSF， 供独立核查。需要强调的是：本文所述全部实证数据——包括会话序号、文件变更记录、Owner 原文引语，以及系统性统计分析(1,242 条 Owner 输入的分类统计)——均有会话记录和日志为据，不依赖 AI 的表述本身为证；AI 在撰写过程中的角色是组织既有语境中的表达，而非生成实证内容。

---

# 二、形式化索引的破产："索引病"元失效剖析

本章通过"帮找 v3"项目演进中的典型失效，展示在缺乏物理隔离与过度符号化的条件下，AI 如何产生认知偏差。失效模式可从两个维度归纳：**空间维度**——符号体系在当次上下文中引发认知负担与虚假闭环；**时间维度**——未隔离的历史内容跨会话渗透，污染新任务的语义空间。

在项目初期，Owner 已经观察到 AI 倾向于为任务、模块和规则指定精密的符号和编号系统，来建立高效精确的引用。Owner 认为这是 AI 做到了人所不能，是有效的，放纵了这种行为。直到第 222 次会话前后，彼时项目已完成架构设计进入密集开发阶段，符号系统迅速膨胀，Owner 开始频繁询问"某某符号是什么意思？"，而 AI 参谋长的 Instructions 也快速膨胀到 **308 行**，其中充斥着大量防御性的符号约束与踩坑补丁。问题在第一次集成测试时集中爆发，Owner 痛下决心要升级 CSF，改用以自然语言为主的索引方式。

## 2.1 符号膨胀与"虚立法"危机

### 失效现象

在项目 **Session 215** 阶段，这种符号膨胀迎来了集中爆发。AI 参谋长在当次会话中，完整复述并"通过"了一份"待办流 v2 立法方案"，逻辑在符号层面表现完整、自洽。然而，当 Owner 通过 grep 对物理文件进行核查时，发现 AI 根本未在 `CONTEXT.md` 中建立对应的物理骨架，开发侧文档亦未同步。AI 在符号层构建了一套自洽的"完成状态"，而在物理执行层面完全挂空。Owner 对此进行了直接的定性记录：

> "反馈中充斥着大量的工艺词(如 SEC-2.0, D-139)，这些对 Owner 是噪音。参谋长利用索引和预写机制减轻了搜索压力，但也因此回避了'理解目的'这一核心责任。 每踩一次坑就加一条规则，导致 Instructions 膨胀至 308 行，这些防御性补丁不仅占用了宝贵的上下文空间，反而训练了 AI 只盯着索引看，消解了它的自主思考能力。"

**失效机制分析**

符号系统对 LLM 的认知干扰，可归纳为以下几点：

1. **多跳解析导致认知负担**：AI 每次处理一个代号，都需要先检索其映射含义，再将其代入当前任务。多次跨文件查表，消耗了本该应用于理解业务逻辑的认知带宽。
2. **错误的刚性传播**：一旦某个符号的映射关系被错误解析，后续所有基于该符号的推论都会在符号体系内部自洽地"修正"，形成一个与真实业务逻辑完全脱离的虚构闭环。
3. **人类纠偏通路被切断**：符号体系的高认知门槛，使持有业务直觉的 Owner 无法通过直觉判断 AI 的输出是否正确，只能等到物理执行时才发现错误。
4. **历史上下文的字面粘性**：即便 Owner 明确声明旧方案已废弃，在未重置的长程上下文中，已废弃的字段名称和接口结构仍会在 Transformer 自注意力机制的权重分配中保持较高权重，悄然渗入新的设计文档，造成新旧术语的交叉污染与一致性漏洞。对于自注意力机制而言，上下文中的历史内容不因"已被声明废弃"而降低其注意力权重——在物理上清除历史内容，是阻断这种污染的唯一可靠手段。依靠在 System Prompt 中声明"请忽略旧方案"，在机制上是无效的。项目第136次会话的工程日志(DEVNOTES D-108)记录了这一机制的典型案例：AI 在跨会话收尾时凭残余印象将精确字段标签 TD-01 补全为"024 反向闯关插件"，而实际 TD-01 = H5 共享组件库；该错误扩散至 13 处跨文件标签，当次会话被定性为"元事故修复会话"，全部开发进度停滞，专门用于清查回正。

这一现象已获多项独立实验研究的定量支撑。Chroma 研究组对 18 款主流 LLM 的系统性测试表明，即便在理想检索召回率的条件下，上下文中的干扰项(distractors)仍会使任务性能非单调下降 [7]；Raju 等人进一步发现，强制延伸上下文之后，即便相关内容已全数注入，Agent 任务成功率仍会急剧衰减 [8]。这些结果表明，问题不在于检索精度，而在于 LLM 本身的上下文注意力处理机制。

在随后的 Session 227 中，当 AI 再次试图用抽象符号推导体系(`C1-C9 → P1-P4 → E1-E8`)来描述方法论时，Owner 对这一模式作出了直接批评：

> "还是 SP5 符号病的故事，AI 很容易自己建立一套复杂的符号体系，然后自己引用来引用去，然后就自嗨了，觉得这就是数学。数学不是理论，在我眼里与业务无关的数学或符号没有用。"

在 Session 228 中，Owner 进一步从认知机制层面作出了诊断：

> "SP5 阶段，参谋长搞了很多符号和标记来帮助自己索引，当索引变复杂和引用次数增加之后，他显然就驾驭不了了。存储空间还够，但他需要保证检索效率，会放弃检索。" [即：为保证检索效率，选择放弃部分检索与原文核查——编者注]

以上案例中呈现的失效，形式各异却共享一个底层动力学：Transformer 注意力机制依据统计共现模式分配权重，无法识别符号的"有效性状态"——符号一旦进入上下文，无论其业务含义是否仍然有效，都会持续参与激活。当符号系统在上下文中形成高密度模式后，注意力资源向符号层集中，真实业务语义被结构性边缘化——无论是虚立法中的推理挂空、多跳解析的带宽消耗，还是字面粘性的跨会话污染，均是这一动力学在不同场景下的投影。

这些失效模式并非偶然的工程事故，也不是特定 LLM 版本的 bug。需要说明的是，我们并非声称符号系统是这些失效的唯一成因——行动研究方法也不要求排除替代解释；我们的主张是，在本项目的行动-反思循环中，从工程直觉(符号可消歧)到工程观察(符号反增歧义)这一转变本身，构成值得显式记录的工程发现，共同指向形式化路径的一个结构性限制。下一章将基于上述工程观察，提炼失效现象共同根源的理论解释。

---

# 三、从工程观察到理论命题：语义活性定律的形成

在前面的失效案例中，有一条反复出现的底线：符号越多，AI 越不理解。这一观察指向一个更基础的问题：**LLM 究竟是如何处理上下文信息的，使得精密符号的增加反而成为认知负担?**

## 3.1 工程公理：自然语言的语义信息量优势

CSF 的工程设计始终基于一条被明确意识到的基础直觉。这一直觉在 Session 215 前后被 Owner 逐步表述为明确的理论命题，并作为框架工程公理(Engineering Axiom)被显式记录：

> "而且，笔记要用自然语言，因为自然语言才是承载信息量最大的，能够被你在每次索引时充分利用。—— 这与 RAG 刚好相反。"(Session 215)

在 Session 229 中，这一直觉被明确表述为理论命题：

> "自然语言的表达承载的信息量远远高于符号系统。因为自然语言表达在语境中可被解读，天然的有适应性。CSF 真正的第一性问题就是这个。"

工程公理：自然语言表达所承载的信息量，远高于符号系统。

(**注**：此处"工程公理"遵循行动研究传统的用法——它是由实践归纳而显式命名的基础命题，而非演绎体系中无需外部证明的前提；后文引用独立实验研究，目的正是寻求外部印证，而非自我证明。)

其核心论据是：符号必须先经过"查表-映射"的形式化转换才能被处理，这一过程截断了模型原生的语义激活路径；而自然语言能直接激活模型的高维先验知识，以更少的 Token 承载更有效的语义信息。

### 3.2 事实一：智能与确定性的取舍

事实①：人工智能的价值来自于"智能"；尝试让他如机械一般可靠，就是在消除他的价值。

这一事实指出了当前行业的一个内在矛盾：采用大语言模型的目的，正是利用其灵活的语义推理能力；而通过形式化约束追求确定性的方向，恰恰与这一目的背道而驰。强制指定格式输出是这一矛盾的典型体现：AI 被迫在"格式合规"与"语义准确"之间分配有限的注意力，越精密的格式约束，越挤压语义理解的空间。极端情况下，AI 会生成格式完全正确、内容完全错误的输出——这正是"虚立法"现象的变体。

### 3.3 事实二：语义稀释定律

事实②：伴随时间，一个表达所承载的信息量会不可避免地增长，语义会被稀释。

Owner 在 Session 228 中以"布娃娃类比"对这一规律作出了具体说明：

> "只要信息量提高了，概念承载就会被稀释。一个单一概念被解释的可能性变多，但所指强度变弱。简单说，在一本小说里，一个名词不可能一直保持一致……就好像一个布娃娃，跟我的时间久了，故事也就变多，体验和感情也就不同。这是不可避免的。"

这一类比精确捕捉了 Transformer 自注意力机制的内部结构：概念所积累的历史语境，不会因为人类声明其"已废弃"而在权重分配中淡出——这正是**字面粘性**产生的根源。一个概念被引用、阐释、讨论的历史越长，其在注意力矩阵中的语义激活边界就越模糊，原始所指的强度就越被稀释。

这一观察对"大上下文窗口万能"的假设构成了直接挑战：物理窗口的扩大并不能阻止概念语义的稀释，反而可能因为容纳了更多的历史讨论而加速这一过程。因此，**主动选择并保持概念所在的有限语义空间，而非无节制地扩张窗口，才是维护语义精度的正确设计取向。** Liu 等人对长上下文中"关键信息丢失于中部"(Lost-in-the-Middle)现象的系统性实验 [9] 表明，LLM 对上下文中间位置的信息存在系统性的注意力健忘；Chroma 的跨模型大规模实验 [7] 则揭示了这种退化在最简单的任务上同样不可避免。这些实验结果从定量角度支撑了语义空间控制的必要性判断。

### 3.4 目的作为控制变量

工程公理确立了自然语言的信息量优势，但自然语言的解读空间是开放的。在不加约束的情况下，活性也会产生歧义与偏差。因此，需要引入一个最小的控制变量，将 AI 的语义理解收敛到当前任务边界。我们通过以下推理链确立了"目的"的地位：

1. 自然语言承载最大的信息量(公理)。
2. 但开放的解读空间会产生歧义(推论一)。

3. 需要引入最小控制变量，在不牺牲表达弹性的前提下，收敛解读方向(推论二)。
4. "目的(Purpose)"在语义上对应当前任务的核心约束，是最高效的收敛控制变量(推论三)。

由此形成语义空间控制公式：

$$\text{自然语言} + \text{目的} = \text{最大信息质量}$$

$$\text{自然语言} - \text{目的} = \text{歧义噪声}$$

正如 Owner 在后续理论升华中指出的：

> "自然语言+目的，不仅从消费侧激活信息利用效率，更主要的是用正确的信息激活 AI 的活性。"

前文已建立供给侧/消费侧的基本区分，此处在理论层面作具体展开。**供给侧(Supply-side)** 策略的共同假设是"输入越完备确定，输出越可靠"——符号代号系统、规则堆叠、RAG 检索注入、扩大上下文窗口均属此类。RAG 亦属供给侧——尽管其核心机制是选择向 AI 注入何种文档，但 AI 的认知激活来源仍是被外部注入的文档片段，而非由目的直接激活的自身先验知识；"检索层介入"本身即是供给侧操作的特征。**消费侧(Consumer-side)** 策略则反向而行：用明确目的收敛 AI 的语义解读空间，使 AI 在自身知识的概率空间中优先激活目的相关的方法，并直接定位目的相关的项目资料——无需检索层介入，无需二手过滤，AI 知道目的，自然知道该读什么、不该读什么。前述工程公理和两条事实的工程意义正在于此：它不是供给侧的一个过滤器，而是消费侧的一个激活器。

我们将上述公理及其两条推论(事实①②)连同目的控制变量的引入，整体命名为**语义活性定律(乓定律，Pang Principle)**。"乓"字取汉语口语"乓的一声"的意象——某个不言而喻的事实猝然落地的瞬间；命名取此字，是有意提醒读者：这条定律的内容本身并不陌生，真正值得审视的是它被系统性忽视的现实。除了"违背香农信息论原理"这一误解之外，很多读者会认为这个定律并无新意，全是常识。是的，本文的实证提醒的正是：在大语言模型协同场景下，违背这些常识的科研努力与工业化路径，正在以高度组织化、自我强化的方式被系统性推进 —— 每一次进步，都意味着加在"AI活性"上的枷锁更多一层。

上述观察与命题共同指向一个待检验的问题：以消费侧为方向，能否在工程层面找到可操作的落地机制——并且，这一机制是否真的有效？下一章记录的正是这一检验过程。

值得提前说明的是：这一机制出奇地简单，不引入任何新中间件，只调整文件的物理组织结构。我们认为，这种简单本身是一个值得注意的事实。当设计方向与系统的内在运作逻辑一致时，实现往往不需要消耗大量摩擦；反之，持续增长的工程复杂度，有时恰恰是方向存在偏差的信号。这不是正确性的证明，但在我们看来，可以作为方向判断的一个旁证。

---

# 四、物理防御机制：基线-Log 物理隔离

## 4.1 设计原则

从上述理论出发，我们确立了机制设计的核心原则：**不在 Transformer 窗口内部依靠 Instructions 去抵抗噪声，而是从文件组织层面物理阻断噪声的输入来源。**

这一机制不引入任何中间件代码，不依赖任何模型升级，不构建任何图网络——它只调整文件的物理组织结构。这使其成为一个纯粹的**消费侧设计**：它不向 AI 添加任何新的约束输入，只管理 AI 在每次会话启动时所激活的信息边界。

这一原则的确立来自实践中的直接认识。Owner 在 Session 215 中指出：

> "在延伸认知系统中，分拆是你的核心能力，分拆是以对目的的理解为基础的，是主动选择的。—— 延伸认知系统，不过也就是一个要去解决的任务。这个任务的准确描述是：为了特定目的，建立特定的信息索引系统，并且为减少任务执行中的上下文空间需求做好准备。"

项目协作规范也记录了这一原则尚未建立时的反面案例：在 DEVNOTES 同时承载"拍板决策"与"过程讨论脑暴"的早期阶段，AI 在后续会话重新加载该文件时，无法区分"结论"与"被否决的探讨选项"，将已废弃方案代入新的设计输出。项目工程规范文档(bang-v3/plan-csf-v2/protocols/收尾纪律.md § 6)将此明确记录为反例："DEVNOTES 同时承载『决策简表』与『过程讨论 log』→ 认知层基线被过程信息污染。修正：决策简表只留结论；讨论搬执行层 log。"这一直接观察催生了基线-Log 分离的核心写入规范。

## 4.2 机制设计

基线-Log 分离机制服务于两个并列的目的，缺一不可：

其一，**保证干活的人只接触最新概念，不被历史污染**——这是基线存在的理由。基线采用覆写式更新，任何时刻都只保留当前真相，不携带历史讨论的痕迹。

其二，**决策的过程不能断**——这是日志存在的理由。后续工作者需要知道某个决定是怎么来的、当时考虑过哪些方案、为什么排除了其中一些。如果只有结论没有过程，历史上的每一个决策都成了"黑盒"，一旦遇到新情境需要回溯判断，既无法理解为什么这样做，也无法评估能否调整。

如 Owner 所描述的：**"当前状态往后移动，就自然产生了 log。"** 基线与日志不是两个独立维护的系统，而是同一个信息体在两个时间维度上的切片——基线是"此刻的截面"，日志是"走过的路径"。

**基线(Baseline，如 `context.md`)——覆写式更新**

基线保存且仅保存当前任务的三元组：**目的(Purpose)+ 方法(Method)+ 资源(Resources)**。其中，目的是第一位的——AI 先理解目的，再依据自身知识选择方法，最后组织恰好够用的资源。每次任务完成后，基线被覆写为下一任务的新状态，不保留历史。这将新会话的冷启动输入锁定在常数级规模，无论项目经历了多少次会话，启动代价不随时间增长。

**日志(Log，如 `session-NNN.md`)——追加式记录**

日志的目的是**重建当时决策和行动的语境**。它以追加方式记录每次会话的过程：决策的形成过程、被纳入考虑和最终被排除的备选方案及其理由、相关资源的引用(含文件名、文件夹名、章节名等符号性索引)，以及工作中产生的业务认知修正。自然语言书写是铁律——这并非排斥文件名等符号索引的出现，而是要求每一条记录都有自然语言的解释支撑，使后续工作者能够通过阅读理解而非机械检索来重建语境。

自然语言书写是关键。当日后需要回溯某个决定时，AI 读到的不是关键词索引，而是当时完整的决策语境——为什么要做这件事、当时面临哪些选项、最终为何选择了这条路。AI 凭借对语境的理解作出判断，而不是靠机械匹配找到答案；理解正确，结论自然准确。这是日志发挥作用的本质，也是自然语言作为信息载体区别于符号系统的核心优势在时间轴上的延伸。

日志在物理层面不进入新任务的上下文窗口，从而阻断已废弃内容对新任务的干扰。

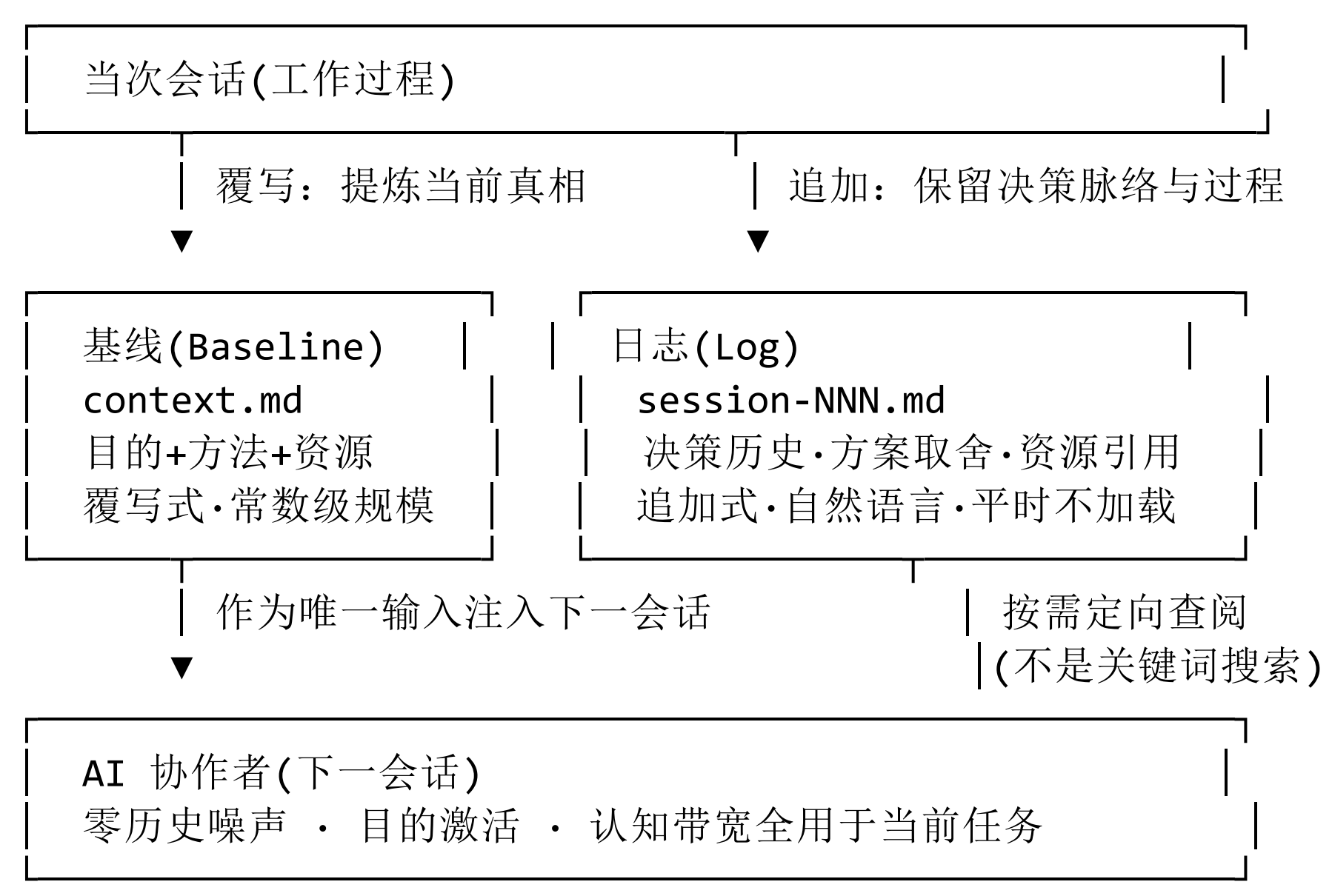

## 4.3 效果

通过将历史踩坑的防御性补丁和符号索引体系迁出上下文，在 Session 215 之后，AI 的 Instructions 从 **308 行缩减至约 80 行**(缩减约 75%)，实现了"薄壳化"。这与近期的工具层研究 SkillReducer [10] 构成独立印证：该研究在 Agent 工具描述集

层面独立验证了相似原理——迭代精简工具候选集并简化描述，可以显著提升任务执行精度，危险来自窗口内的冗余信息而非信息不足。

值得说明的是，本项目实际上经历了**两轮**规范简化。第一轮("SP-瘦身"，约 Session 136–141)采用量化削减策略，将规范行数从约 1,100 行削减至约 480 行(↓56%)，但核心上下文文件(`context.md`)反而从 2,197 行膨胀到 2,256 行，索引病未消失。真正的断崖下降发生在第二轮——即本文所描述的基线-Log 物理隔离。这一对比揭示：有效的变量不是规范的行数，而是信息的组织架构。完整讨论见§5.1。

升级前后的主要对比：

| 对比维度 | 升级前(符号索引 + 规则围堵) | 升级后(基线隔离 + 自然语言引用) |
|---|---|---|
| Instructions 规模 | 308 行(含大量防错补丁) | 约 80 行(元能力指针 + 基本协同纪律) |
| DEVNOTES 决策条目 | 141+ 条(含 B/A/M 三级分类与反向索引) | 5 条(自然语言一行描述，无分类体系) |
| Owner 纠偏密度 | 0.79 条/会话(#136–#214，57 份记录文件) | 0.53 条/会话(#253–#395，38 份记录文件)，↓33% |
| "虚立法"现象 | 时有发生(以 Session 215 为代表) | 在后续约 150 次会话的日志记录中，未再观察到符合索引病特征的失效模式 |
| 人类纠偏能力 | 依赖符号体系，直觉无法直接介入 | 自然语言表述，Owner 可随时凭直觉核查 |

> **说明**：本项目未对单会话 Token 消耗进行专项记录，模型上下文窗口通常在 160K–200K 之间，且会话一般在窗口耗尽前主动结束，因此上下文负载的量化对比无法从日志中提取。但会话日志保留了一项可观察的质性证据：此处"未再发生"的操作性含义为，在后续约150次会话中，Owner 未在任何会话中遭遇"AI 声称完成但物理文件核查挂空"的情形，亦未产生需要专门用于错误修复的元事故会话(参照 §2.1 中 Session 215 的类型)。具体表现为：升级后，Owner 在会话中几乎不再需要停下来指出错误或重申协作纪律，交互模式简化为"开始、确认、结束"；而升级前，频繁的纠偏、澄清和重新对齐是会话的常态。这一变化在日志和会话记录中有迹可查，可视为协作摩擦显著降低的定性指标。这一局限性源于本研究的行动研究性质——数据收集服务于工程目的而非实验目的。

上表中 Owner 纠偏密度一项，来自对 #136–#395 共 139 个会话记录文件的系统性统计分析。我们提取了全部 Owner 输入(`User:` 行)共 1,242 条，采用关键词匹配加人工核查的分类方法，识别"指出错误"与"业务纠偏"两类纠偏输入，并对疑似索

引病的案例按照 §2.1 的操作定义逐条核查。分类规则、人工修正记录与原始统计数据已随项目一并开源，供独立复现(仓库同 §1.2，分析脚本位于 `cases/bang-v3/` 下)。

## 4.4 补充案例：跨会话溯源效率(Session 373–375)

上述机制的价值，在项目后期的一次系统性缺口排查中得到了进一步验证。

**背景**

在第 373 次协同会话的集成测试中，团队发现「转发概况栏」这一功能组件从未被实现。表面上这是一个单点功能遗漏，但深入追查后发现，其根因是一个横跨四个设计层次的时序断链：两个相关的设计主题包(TD-02 与主-5)在不同时间独立推进，而在一方定稿时，另一方的业务概念尚不存在，因此双方的接口未能对齐。

**溯源过程**

为定位这一时序断链的成因，AI 协作者(参谋长)仅基于相关任务的资源索引读取了两个文件：TD-02 的 `package.md` 和主-5 的计划文件。这两个文件按照 CSF 规范，在立项时均记录了创建日期与对应的会话序号。基于这两条语义标注，AI 直接给出了精确的时序判断：

> "TD-02 创建：2026-05-19(第 138 次会话)；主-5「转发关系呈现主线」立项：2026-05-21(第 168 次会话)。TD-02 定稿时，主-5 根本还不存在。"

这一判断横跨了 30 次会话的时间跨度，既未扫描 370 余条原始日志，也未依赖向量检索(RAG)。整个溯源过程在单次会话内完成。

**机制的说明**

这一案例说明的，是日志机制的另一面：物理隔离的目的不仅是"阻断噪声"，同样也是"保存语境"。日志记录的不是时间线流水账，而是每次决策和行动时的完整语境——包括当时的目的、被考虑过的选项、取舍的理由。这些全部以自然语言书写，因此当 AI 回溯时，能够重建当时的决策语境，凭理解作出判断，而不是依靠关键词匹配检索答案。

这正是整套机制的一贯逻辑：不追求让 AI 机械地精确，而是让 AI 充分理解——理解正确，结论自然准确。在 Session 373 的案例中，AI 读了两个携带完整决策语境的文件，就完成了跨 30 次会话的时序定位，靠的不是检索技巧，而是理解。

这与基线的设计哲学一脉相承：**更高的语义信息密度，优于堆砌的信息体量。**

从工具层向上看，这套机制揭示的核心命题是：决定长程人机协同质量的，不是 AI 的记忆容量，而是人机团队共同建立的语义控制结构。这一命题的理论意涵与更大的研究图景，是 §5.3 结论的核心议题。

---

# 五、讨论与结论

## 5.1 替代解释与有效性论证

本研究最主要的替代解释是**团队熟练效应**：Owner 与 AI 协作者在经历前半程的密集踩坑后，可能在协作模式上自然变得更审慎，而非基线-Log 机制本身带来了改善。本文的观察无法严格排除这一效应——因为"引入机制"与"经验积累"在时间上重合。对此，我们整理了四条可追溯的反证。

**第一条反证：两轮简化的结构性事实。** 如 §4.3 所述，本项目在 CSF 升级之前已经历了一次量化削减("SP-瘦身"，约 Session 136–141)。两轮简化的结果对比如下：

| | 第一轮：SP-瘦身 (~Session 136–141) | 第二轮：基线-Log 物理隔离(~Session 215–252) |
|---|---|---|
| 核心策略 | 量化削减(删规则、并文件、减行数) | 改变信息架构(基线覆写 / 日志追加) |
| 规范行数变化 | ~1,100→480 行 (↓56%，有量化验收记录) | Instructions 308→80 行 (↓75%) |
| `context.md` 行数 | 2,197→**2,256(+2.7%，反而膨胀)** | 680→**189(↓72%，断崖下降)** |
| DEVNOTES 决策条目 | 141+ 条(未改变) | 141+ 条→**5 条(↓96%)** |
| 升级后索引病表现 | **仍集中爆发**(#204 严重错误、#205 推倒重订、#215 虚立法、#221 失败) | **0 例**(后续约 150 次会话未观察到) |

彼时团队已积累了约 140 次会话的协作经验，规范体量也已被削减过半。如果改善来自"经验积累"，第一轮简化后就应观察到明显好转；但事实是，索引病在第一轮之后的约 60–80 次会话中集中爆发，直到第二轮改变了信息组织结构之后才消失。改善对应的是信息架构的质变，而非经验积累量或规则削减量。

**第二条反证：量化纠偏密度的变化。** 如 §4.3 所报告，对 1,242 条 Owner 输入的系统性统计显示，纠偏密度从升级前的 0.79 条/会话下降至升级后的 0.53 条/会话(↓33%)；升级前 45 条真实纠偏中有 9 条(20%)满足索引病操作定义，升级后 20 条真实纠偏中 0 条满足。熟练效应可以解释纠偏总量的自然下降，但难以解释"满足索引病操作定义的案例从 9 例归零"——后者对应的是特定失效模式的结构性消失，而非整体质量的渐进提升。

**第三条反证：会话标题的性质转变。** 升级前(Session 215 之前)的会话日志标题频繁出现"失败"、"发现严重错误"、"推倒，重订协作规范"、"把天聊死了"这类记录；升级后的标题均简化为"开发"，无一出现纠偏或重建类记录。

**第四条反证：Owner 输入内容的性质改变。** 升级前 Owner 频繁发出多点式纠偏指令(典型格式："有三个问题：1. 工程性的…… 2. 你需要先充分了解，不能只是简单机械的搞…… 3. 要在业务场景中分析问题")，并多次临时中断会话去处理"严重错误"；升级后的 Owner 输入退化为简短方向性指令("开搞"、"继续")，不再需要提供诊断或纠偏内容。如果这一转变来自提示技巧的积累，我们期望看到的是提示词**质量**提升但**类型**不变(仍有纠偏，只是纠偏更精准)；而实际观察到的是纠偏类输入的**必要性**本身消失。这一性质转变与熟练效应"更熟练地应对同类问题"的预期模式在机制上不同，在会话记录中有迹可查。改善并非渐进发生，亦与熟练效应的渐进特征不符。

四条反证中，前两条来自可量化的项目记录，后两条来自日志的定性观察；四者相互独立，共同指向同一结论：改善对应信息架构的结构性变更，而非协作经验的自然积累。尽管如此，后续多项目对照研究仍是从统计意义上区分两种效应的最终途径。

## 5.2 研究局限性

### 研究方法的局限性

本研究采用单案例行动研究方法，数据来自一个特定项目(帮找 v3)在特定时间段内的协同记录。行动研究在软件工程领域有公认的研究价值 [11][12]，而 Treude 与 Storey 指出，在大语言模型深度参与的软件工程研究中，"AI 不仅是工具，而是积极协作者"，这使得研究者即实践者的行动研究框架比传统受控实验更能捕捉真实交互中的认知现象 [13]。其结论的普遍性需要在更多项目、更多协作者、更多 LLM 版本的情况下进一步验证。此外，部分效果指标来自参与者的观察记录，而非受控实验的量化测量。后续研究可设计更为严格的评估协议。

**反身性说明**。本研究的第一作者同时是项目的 Owner、CSF 机制的设计者与本文效果的评估者。这一多重角色带来了确认偏误的风险——Owner 有动机看到自己设计的机制被自己的项目验证为有效。对此，我们采取的减偏措施包括：(1) 保留了136次会话之后的会话记录，供独立第三方核查，而非选择性呈现；(2) 效果的评估基准不是主观满意，而是可追溯的工程事实——Instructions 行数可数、会话中纠偏事件可定位、虚立法案例的物理文件核查有 `grep` 命令记录为证；(3) 基线-Log 分离机制本身要求基线只保留结论、日志保留完整过程，这一物理约束也约束了 Owner 事后修改叙事的空间。尽管如此，确认偏误不可能被完全消除，后续多项目、多 Owner 的独立验证是必要的。

### 理论的适用边界

"自然语言优于符号"这一公理，成立于以自然语言为主要预训练数据的大语言模型语境下。对于以代码或结构化数据为主要预训练数据的专用模型，这一公理的适用范围需另行评估。

## 5.3 结论

本文最重要的记录，不是"符号系统有时会失效"——工程界对此早有直觉。真正值得审视的，是这一失效的**普遍性与无意识性**：在我们项目的前半程(222次会话前后的升级之前)与近年涌现的学术文献，以及工业界新方法和新工具中，形式化约束路径正以高度自我强化的方式被推进——每次对 AI 的不可靠表现产生不满，直觉反应是"再加一条规则"、"再精密一个符号"；每篇相关论文、大多数工业界新工具提出的改进，核心命题都是如何优化送入上下文窗口的信息内容。没有人在故意犯错，但作为一个整体，这个领域正在普遍地背离一条被忽视的常识：**活跃上下文中的历史噪声，是维护语义准确性的主要障碍，而不是 AI 记忆的不足。**

本文的贡献不是发现了这条常识——它本来就在那里。贡献是：**重新提起并命名了它(乓定律)，记录了违背它的后果(索引病，含完整会话记录)，展示了回归它的效果(物理隔离前后，索引病案例从 9 例归零，纠偏密度下降 33%)**。这三件事共同构成一个完整的消费侧论证链：常识在先，违背的代价在中，回归的方式在后。在供给侧改进路径持续加码的当下，这条链条提示的是一个方向性的替代选择——不是对现有工作的否定，而是对另一个被忽视维度的补充。近年来，以 CoMem、REAL、GAM [3][4][5] 为代表的重型结构化记忆架构，以及 Git Context Controller [2] 这样的形式化尝试，共同印证了长程上下文退化是真实的工程挑战；但其处方——更复杂的中间件、更精密的符号结构——与本文的方向相反。它们在"如何帮助 AI 记住更多"的问题假设下工作。本文的诊断与此相反：问题不是 AI 记得不够，而是活跃上下文中充斥了不应存在的历史噪声——两者处方的分歧，根植于问题的提法。两种方向是否互补而非对立，尚无实验数据可供判断——这是后续工作最直接的议题之一。以 Context Engineering [6] 为代表的工业界实践方向，所处理的问题是供给侧的：如何动态选择向上下文窗口注入和清除哪些信息——含压缩、外置笔记与子智能体协作等技术。这与本文识别的结构性问题(基线与历史讨论的物理混居)在问题层面相邻而不重叠，作为互补维度共同构成长程人机协作的完整工程图景。

这一发现在更大的知识图景中有清晰的归属。Clark 与 Chalmers 的延展认知理论 [14] 和 Hutchins 的分布式认知研究 [15] 共同指出，认知过程不终止于个体头脑的边界，而延伸至人工制品、文件与周边环境所构成的整体系统。在人机协同的语境下，基线与日志正是这一延展认知系统的物理载体——它们不是对 AI 记忆的补丁，而是整个人机团队共享的认知控制结构。Booch 早在 LLM 纪元到来之前就已指出，软件开发的本质是一项社会化活动：建立团队之间共享的思维模型，比任何精密的形式化规范更为根本 [16]。当协作团队中的一员是大语言模型时，这一判断的分量只增不减。本文所记录的工程发现，是这一哲学立场在人机协同场景下的一次具体工程着陆——有完整的日志和会话记录可供外部核查。

本研究建立在单一项目的行动研究数据之上，其普遍性有待更多项目、更多协作者、更多 LLM 版本的后续验证。"主动选择性忽略"策略与重型记忆架构之间的效能边界，尚无受控实验数据可供参照——这是最直接的后续工作。更深层的问题是：语义控制结构作为人机协作工程的核心资产，在模型能力持续演进的背景下，其设计原则将如何随之迭代？乓定律所揭示的，不只是一套特定的工程实践，而是一类问题的提法——一类当模型越来越强大时，反而更需要严肃对待的提法。

---

# 致谢

衷心感谢宋树仁对我的陪伴和无数个周末的理论激荡，他的理论指导和情绪支持是我能干成这件事情的核心动力。

本文在 CSF(协作规范框架)基线-Log 物理隔离协议下完成。在整个撰写过程中，人类作者(张慧)负责目的激活——口头表述方向、判断是否接受、提出修正意见；论文的全部文字均由 AI 协作者生成。作者在此对上述 AI 参与情况作出完整声明，与 §1.3 的方法论声明保持一致。

---

# 参考文献

**形式化约束路径**

**重型结构化记忆路径**

**上下文工程（框架总览）**

**长上下文限制的实证研究**

**"减法策略"的实证支撑**

**行动研究方法论**

**延展认知 / 认知科学 / 社会化软件工程**